# All-optical retrieval of the global phase for two-dimensional Fourier-transform spectroscopy

**Alan D. Bristow, Denis Karaiskaj, Xingcan Dai and Steven T. Cundiff***

*JILA, University of Colorado and National Institute of Standards and Technology,
Boulder, Colorado 80309-0440 USA*
*Corresponding author: cundiffs@jila.colorado.edu*

**Abstract:** A combination of spatial interference patterns and spectral interferometry are used to find the global phase for non-collinear two-dimensional Fourier-transform (2DFT) spectra. Results are compared with those using the spectrally resolved transient absorption (STRA) method to find the global phase when excitation is with co-linear polarization. Additionally cross-linear polarized 2DFT spectra are correctly "*phased*" using the all-optical technique, where the SRTA is not applicable.

---

## References and links

1. S. Mukamel, "Multidimensional femtosecond correlation spectroscopies of electronic and vibrational excitations," Annu. Rev. Phys. Chem. **51**, 691-729 (2000).
2. D. M. Jonas, "Two-dimensional femtosecond spectroscopy," Annu. Rev. Phys. Chem. **54**, 425-463 (2003).
3. M. Cho, "Coherent two-dimensional optical spectroscopy," Chem. Rev. **108**, 1331-1418 (2008).
4. R. R. Ernst, G. Bodenhausen and A. Wokaun, *Principles of nuclear magnetic resonance in one and two dimensions* (Oxford, 1987).
5. P. Hamm, M. H. Lim and R. M. Hochstrasser, "Structure of the Amide I band of peptides measured by femtosecond nonlinear-infrared spectroscopy," J. Phys. Chem. B **102**, 6123-6138 (1998).
6. J. D. Hybl, A. W. Albrecht, S. M. Gallagher Faeder and D. M. Jonas, "Two-dimensional electronic spectroscopy," Chem. Phys. Lett. **297**, 307-313 (1998).
7. M. C. Asplund, M. T. Zanni and R. M. Hochstrasser, "Two-dimensional infrared spectroscopy of peptides by phase-controlled femtosecond vibrational photon echoes," Proc. Natl. Sci. USA **97**, 8219-8224 (2000).
8. O. Golonzka, M. Khalil, N. Demirdöven, and A. Tokmakoff, "Vibrational anharmonicities revealed by coherent two-dimensional infrared spectroscopy," Phys. Rev. Lett. **86**, 2154-2157 (2001).
9. L. Lepetit, G. Chériaux and M. Joffre, "Linear techniques of phase measurement by femtosecond spectral interferometry for applications in spectroscopy," J. Opt. Soc. Am. B **12**, 2467-2474 (1995).
10. L. Lepetit and M. Joffre, "Two-dimensional nonlinear optics using Fourier-transform spectral interferometry," Opt. Lett. **21**, 564-566 (1996).
11. J. D. Hybl, A. W. Albrecht Ferro and D. M. Jonas, "Two-dimensional Fourier transform electronic spectroscopy," J. Chem. Phys. **115**, 6606-6622 (2001).
12. N. Belabas and M. Joffre, "Visible-infrared two-dimensional Fourier-transform spectroscopy," Opt. Lett. **27**, 2043-2045 (2003).
13. T. Zhang, C. N. Borca, X. Li and S. T. Cundiff, "Optical two-dimensional Fourier transform spectroscopy with active interferometric stabilization," Opt. Express **13**, 7432-7441 (2005).
14. M. L. Cowan, J. P. Ogilvie and R. J. D. Miller, "Two-dimensional spectroscopy using diffractive optics based phase-locked photon echoes," Chem. Phys. Lett. **386**, 184-189 (2004).
15. T. Brixner, I. V. Stiopkin and G. R. Fleming, "Tunable two-dimensional femtosecond spectroscopy," Opt. Lett. **29**, 884-886 (2004).
16. P. Tian, D. Keusters, Y. Suzaki and W. S. Warren, "Femtosecond phase-coherent two-dimensional spectroscopy," Science **300**, 1553-1555 (2003).
17. E. M. Grumstrup, S. Shim, M. A. Montgomery, N. H. Damrauer, and M. T. Zanni, "Facile collection of two-dimensional electronic spectra using femtosecond pulse-shaping technology," Opt. Express **15**, 16681-16689 (2007).
18. S.-H. Shim, D. B. Strasfeld, Y. L. Ling and M. T. Zanni, "Automated 2D IR spectroscopy using a mid-IR pulse shaper and applications of this technology to the human islet amyloid polypeptide, Proc. Natl. Sci. USA **104**, 14197-14202 (2007).

19. J. C. Vaughan, T. Hornung, K. W. Stone and K. A. Nelson, "Coherently controlled ultrafast four-wave mixing spectroscopy, J. Phys. Chem. A **111**, 4873-4883 (2007).
20. V. Volkov, R. Schanz and P. Hamm, "Active phase stabilization in Fourier-transform two-dimensional infrared spectroscopy," Opt. Lett. **30**, 2010-2012 (2005).
21. D. Keusters, H.-S. Tan and W. S. Warren, "Role of phase and direction in two-dimensional optical spectroscopy," J. Phys. Chem. A **103**, 10369-10380 (1999).
22. M. Khalil, N. Demirdoven, and A. Tokmakoff, "Obtaining absorptive line shapes in two-dimensional infrared vibrational correlation spectra," Phys. Rev. Lett. **90**, 047401 (2003).
23. X. Li. T. Zhang, C. N. Borca and S. T. Cundiff, "Many-body interactions in semiconductors probed by optical two-dimensional Fourier transform spectroscopy," Phys. Rev. Lett. **96**, 057406 (2006).
24. T. Zhang, I. Kuznetsova, T. Meier, X. Li, R. P. Mirin, P. Thomas and S. T. Cundiff, "Polarization-dependent optical 2D Fourier transform spectroscopy of semiconductors," Proc. Natl. Sci. USA **104**, 14227-14232 (2007).
25. L. Yang and S. Mukamel, "Two-dimensional correlation spectroscopy of two-exciton resonances in semiconductor quantum-wells," Phys. Rev. Lett. **100**, 057402 (2008).
26. L. Yang, I. V. Schweigert, S. T. Cundiff and S. Mukamel, "Two-dimensional optical spectroscopy of excitons in semiconductor quantum-wells: Louiville-space pathway analysis," Phys. Rev. B **75**, 125302 (2007).
27. A. D. Bristow, D. Karaiskaj, X. Dai, T. Zhang, C. F. Carlsson, K. R. Hagan, R. Jimenez and S. T. Cundiff, *to be published*.
28. I. Kutnetsova, T. Meier, P. Thomas, *private communication*.
29. E. H. G. Backus, S. Garrett-Roe, and P. Hamm, "On the phasing problem of heterodyne-detected two-dimensional infrared spectroscopy," *accepted for publication in Opt. Lett.*.

## 1. Introduction

Over the last decade, two-dimensional Fourier-transform (2DFT) spectroscopy at optical wavelengths has proven to be a useful spectroscopic tool; see for example [1-3]. Optical 2DFT spectroscopy is analogous to multidimensional nuclear magnetic resonance spectroscopy [4]. In the optical regime, intense laser pulses are used to explore either the vibronic or electronic resonances, depending on the frequency of excitation. 2DFT allows congested spectra to be unfolded, separating populations from coherent interactions to reveal both dynamic and structural information.

Present day 2DFT spectra are generated by taking a numerical Fourier transform of time-domain measurements, such as a heterodyne-detected photon echo. Multiple pulses excite the nonlinear response in a sample, where the photon echo is one specific case of a larger set of transient four-wave mixing (TFWM) signals [5-8]. In these time-domain experiments, the emission can be measured as a function of the pump time delay, or as a function of a gating pulse delay used to temporally resolve the signal. For 2DFT experiments, the TFWM technique is enhanced so that the phase of the signal can be explicitly tracked, while the incident pump and emission time delays are scanned. The time-domain emission can also be spectrally resolved to improve the detection speed; in which case, spectral interferometry (SI) techniques can be employed to measure the complex spectrum [9].

Sub-cycle phase control is required to perform 2DFT experiments, and is harder to achieve as the excitation frequency is increased. An early example of phase control used SI between second harmonic generated at the pump overlap in the sample and a doubled reference beam [10]. The result of which was a real-time measurement of the interferometer's optical path difference, allowing for manual correction [6,11]. Alternatively, co-propagating continuous-wave laser beams can measure the optical path difference, supplying this real-time information to a feedback loop without impeding the scan range of the interferometer arms [12]. Active feedback loops have been shown to stabilize NIR light up to approximately $\lambda/100$ [12,13]. An alternative to the active stabilization is the use of common path optics based on diffractive optic elements [14,15] and pulse shaping [16-19]. Many of these systems typically suffer from short maximum delays between the pump pulses. Recently spatial patterns have been employed to stabilize the phase of the excitation pulses in infrared 2DFT spectra [20].

Collinear [16], partially collinear [17,18] and non-collinear [6-8] 2DFT experiments have had varying success separating phase contributions from the signal and phase offsets associated with the pump beams or heterodyning interferometers. Obtaining this global phase allows extraction of the real and imaginary components of the 2DFT spectrum [21]. The global phase is associated only with the third-order response function of the material, and is independent of the phase of the excitation pulses or the phase of any additional heterodyning pulses. The importance of this has been demonstrated with respect to finding absorptive lineshapes of vibrational correlations in molecules [22] and in the many-body interactions of semiconductor excitons [23,24].

For non-collinear phase-matching 2DFT experiments, correctly "*phasing*" the spectra has been achieved by matching the complex TFWM spectrum to the spectrally resolved transient absorption (SRTA), recorded in an auxiliary yet *in situ* measurement [2]. The method requires using the two-beam, pump-probe geometry to mimic the exact excitation condition used in the 2DFT measurement, a criterion that is especially strict for semiconductor spectroscopy. However, mimicking these conditions exactly can be tricky since the geometry is different. Moreover, at low excitation densities matching the weak SRTA and complex TFWM spectra can lead to significant errors. Furthermore, SRTA cannot be measured to phase certain cases of 2DFT spectra; namely, cross-polarization [24] and some virtual-echo techniques [25].

In this paper, we address acquisition of the global phase for non-collinear 2DFT experiments without relying on a SRTA measurement. Instead, an all-optical measurement is demonstrated, which is based on using spatial interference patterns at the sample position combined with SI measurements. These demonstrations are performed on 2DFT spectra of semiconductor excitons in multiple quantum wells (MQWs). First, the samples are excited by suitably strong, co-linear polarized pump beams, where the SRTA measurement can provide good comparison to the all-optical scheme proposed. Secondly, a cross-linear polarized excitation case is presented where no such comparison is possible.

## 2. Analysis

Electronic 2DFT spectroscopy in the $\chi^{(3)}$ nonlinear optical regime is associated with the third-order nonlinear response function $S(\tau,T,t)$ of a set of resonances. In the measurements presented here, TFWM signals in semiconductor MQWs are generated by three pulses that are arranged in the box geometry (along with a fourth pulse that traces out the TFWM signal path). In the experiment the conjugate pulse impinges the samples first, resulting in a real photon echo that corresponds to the phase-matching condition $k_S = -k_A + k_B + k_C$, and a "*rephasing*" or $S_I(\omega_\tau,T,\omega_t)$ 2DFT projection [26]. In $S_I(\omega_\tau,T,\omega_t)$ measurements the $\tau$ delay (between the first two pulses) is scanned producing an evolving phase. Since the measurement is performed with phase stabilization the global phase can be obtained at the beginning of the scan, i.e. when $\tau = 0$. The analysis for this is presented below.

*2.1 All-optical phase retrieval*

The global phase of the 2DFT spectrum is a result of the third-order polarization and independent of the input pulses or any other optical paths. In the frequency domain, the third-order polarization can be written as

$$\hat{P}^{(3)}(\omega_D) = \iiint \chi^{(3)}(\omega_D;\omega_1,\omega_2,\omega_3)\hat{E}_1(\omega_1)\hat{E}_2(\omega_2)\hat{E}_3(\omega_3)d\omega_1 d\omega_2 d\omega_3, \quad (1)$$

where $\chi^{(3)}(\omega_D;\omega_1,\omega_2,\omega_3)$ is the nonlinear susceptibility. Each term can be written using complex notation, such that $\hat{P}^{(3)}(\omega) = \hat{P}^{(3)}(\omega) + c.c.$, $\chi^{(3)}(\omega) = \hat{\chi}^{(3)}(\omega) + c.c.$ and $\hat{E}(\omega) = \hat{E}(\omega) + c.c.$. All three pump fields are assumed to have the same power spectrum, with different wavevectors, time delays and phases. So the total field can be expressed as

$$\hat{E}(\omega) = \tilde{E}_A(\omega)e^{ik_A \cdot r}e^{-i\omega\tau_A}e^{i\phi_A} + \tilde{E}_B(\omega)e^{ik_B \cdot r}e^{-i\omega\tau_B}e^{i\phi_B} + \tilde{E}_C(\omega)e^{ik_C \cdot r}e^{-i\omega\tau_C}e^{i\phi_C}, \qquad (2)$$

where $\tilde{E}_i(\omega)$, $k_i$, $\tau_i$ and $\phi_i$ are the reduced electric field, wavevector, delay and phase of the $i$ th beam. Now the notation A, B, C is used to relate the pulses to those chosen in the phase-matching condition above. This expression can be inserted into Eq. (1) with due attention to the complex nature of each term. There are many contributions to the third-order polarization for the sample. If only the $S_I$ technique is considered, then the polarization for the $k_S = -k_A + k_B + k_C$ direction becomes:

$$\hat{P}^{(3)}_{-k_A+k_B+k_C}(\omega) \propto \hat{\chi}^{(3)}(\omega_D;\omega_1,\omega_2,\omega_3)|\tilde{E}(\omega)|^2 \tilde{E}(\omega)e^{-i\omega(-\tau_A+\tau_B+\tau_C)}e^{i(-\phi_A+\phi_B+\phi_C)}. \qquad (3)$$

The TFWM signal can be written as $\hat{E}_S(\omega) \propto i\hat{P}^{(3)}_{-k_A+k_B+k_C}(\omega)$. SI characterizes $\hat{E}_S$ against a reference pulse that has a power spectrum identical to the pump pulses, yet has different phase and delay: thus $\hat{E}_R(\omega) = \tilde{E}_R(\omega)e^{ik_R \cdot r}e^{-i\omega\tau_R}e^{i\phi_R} + c.c.$, where $k_S = k_R$. The intensity of the SI signal is

$$I = |\hat{E}_S + \hat{E}_R|^2 = |\hat{E}_S|^2 + |\hat{E}_R|^2 + \hat{E}_S\hat{E}_R^* + \hat{E}_S^*\hat{E}_R. \qquad (4)$$

In SI the individual field terms are subtracted and the appropriate interference term is chosen based on causality arguments [7], leaving only the $\hat{E}_S\hat{E}_R^*$ term that is equivalent to $S_{SR}(\omega)$:

$$S_{SR}(\omega) = \hat{\chi}^{(3)}(\omega_D;\omega_1,\omega_2,\omega_3)|\tilde{E}(\omega)|^4 e^{-i\omega(-\tau_A+\tau_B+\tau_C-\tau_R)}e^{i(-\phi_A+\phi_B+\phi_C-\phi_R)}. \qquad (5)$$

The delay components can be removed through careful measurement and setting them to zero. For the 2DFT measurement $\tau = \tau_A - \tau_B$ is scanned and is not removed from the entire data set. However, as previously mentioned phase offset is only required at $\tau = 0$. Therefore, when measuring the spectral phase of the signal we may be left with phase offsets that displace the signal. These contributions rotate the entire complex 2DFT spectra and are given by

$$\phi_{P^{(3)}} = \phi_{SR} - (-\phi_A + \phi_B + \phi_C - \phi_R). \qquad (6)$$

This equation relates the phase offsets of the incident pump and reference pulses to the TFWM signal measured with SI at $\tau = 0$, arranged to give the phase of the polarization, $\phi_{P^{(3)}}$.

SI gives spectral phase information, from which a phase offset can be extracted and used in Eq. (6). The individual phase offsets of the three incident pumps and reference pulses need to be obtained elsewhere. In the box geometry, the four pulses overlap at the sample position, resulting in a spatial interference pattern that contains information about all four phase offsets. Using a camera, this can be captured as an image that can be interrogated to extract these offset values. The entire 2D image can be processed simultaneously or pair-wise comparisons of the pulses, such as $-\phi_A + \phi_B$ and $\phi_C - \phi_R$, can be performed. For some measurements the reference (or local oscillator) is in fact the tracer pulse and passes through the sample. In such experiments this analysis is sufficient to determine the global phase offset.

Semiconductors are extremely sensitive to the excitation conditions. Therefore, the reference pulse cannot pass through the TFWM excitation spot on the sample, mandating the use of an additional interferometer. In these cases, the reference pulse is not the tracer pulse and the camera image is insufficient to obtain the true signal phase. Consequently, the treatment of global phase retrieval must be extended to link the tracer to the reference of Eq. (6). This can be done with an additional SI measurement between only the tracer and reference. Following the same recipe for SI, Eq. (4) is rewritten replacing $\hat{E}_S$ with $\hat{E}_T$ for the tracer beam, and the interference term is $\hat{E}_T \hat{E}_R^*$:

$$S_{TR}(\omega) = \left|\tilde{E}(\omega)\right|^2 e^{-i\omega(\tau_T + \tau_R)} e^{i(\phi_T - \phi_R)}, \quad (7)$$

where $\tau_T$ and $\phi_T$ are the delay and phase of the tracer beam. In this case the time delays terms are measured explicitly, so that the remaining phase offset terms are

$$\phi_{TR} = \phi_T - \phi_R. \quad (8)$$

Equating Eq. (6) and Eq. (8) gives

$$\phi_{P^{(3)}} = \phi_{SR} - \phi_{CAM} - \phi_{TR}, \quad (9)$$

where $\phi_{CAM} = -\phi_A + \phi_B + \phi_C - \phi_T$, denoted such because it can be captured on a camera where the four beams overlap and interfere.

### 3. Experimental

The pump source is a 160 fs Ti:sapphire laser oscillator, with a repetition rate of 76 MHz and a center wavelength of about 800 nm (~1555 meV). It is split into four identical pulses, arranged in the box geometry by a new ultra-stable set of cascaded interferometers. This apparatus is the subject of a different paper [27]. These interferometers are phase-stabilized to better than λ/100 using active feedback loops that read an error signal from co-propagating HeNe beams; see Zhang *et al* [13]. As shown in Fig. 1(a), three of the input pulses are the pumps for the TFWM and 2DFT measurements, and the fourth is the phase-stabilized tracer (and reference) pulse. All four beams are focused with a 15 cm lens so that they impinge on the sample at the same location inside the cryostat. The generated TFWM signal is diffracted along the phase-matching direction, which is collinear with the tracer pulse. A collimating lens reconstructs the box on the exit side of the cryostat. The reference pulse is split from the tracer before the sample position and focusing lens and is routed around the cryostat. A beam splitter recombines the reference with either the TFWM or tracer before it is coupled into a spectrometer with a thermo-electric cooled CCD attached.

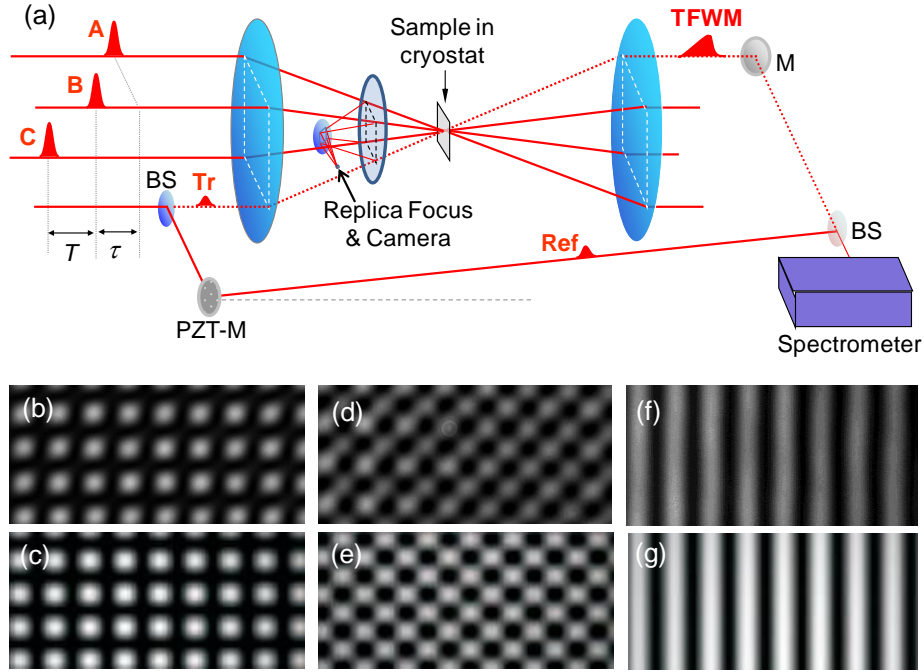

Fig. 1. (a) Schematic diagram of the experimental setup. Pulse A is the conjugate pulse compared to pulses B and C. (Tr = tracer beam, ref = reference beam, BS = beamsplitter, M = mirror and PZT-M = piezoelectric transducer mounted mirror.) Various images of the replica focus are shown in (b), (d) and (f), and are modeled in (c), (e) and (g) for comparison.

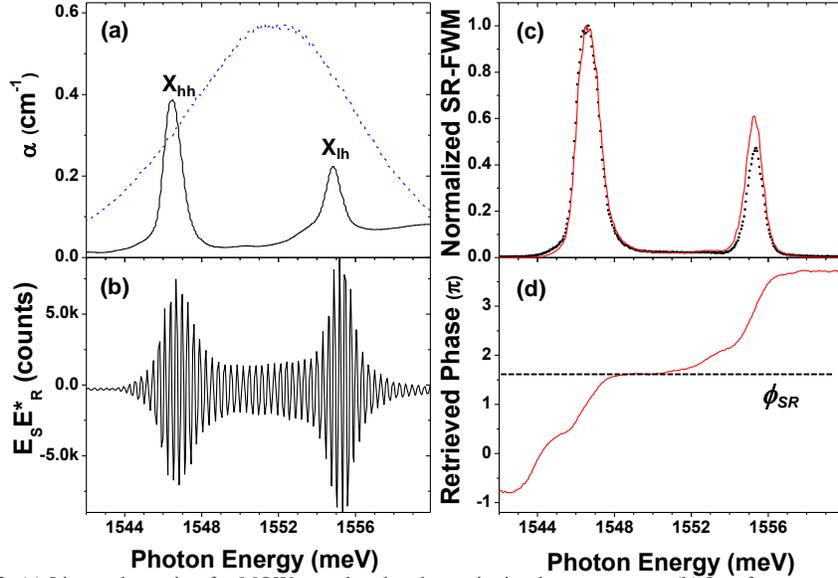

Fig. 2. (a) Linear absorption for MQW sample, plus the excitation laser spectrum. (b) Interference term for the SI between TFWM and reference. (c) Measured (black dots) and retrieved (red line) TFWM spectra. (d) Retrieved spectral phase from the SI.

A replica focus is created away from the main axis of the experiment. This is achieved using a beam sampler, placed between the focusing lens and the sample position, which reflects the converging beams onto a small turning mirror. The beam sampler is aligned to perfectly back reflect the incident beams without the presence of the lens, ensuring identical contributions to the phase due to reflection. The mirror is placed on the axis of the box, so as not to obscure the incoming beams, and reflects them at $45^\circ$ so they focus outside the box. A camera captures the replica focus with a x40 microscope objective magnifying the crossing point. Various camera images are shown in Figs. 1(b), 1(d) and 1(f), corresponding to all pulses (almost) in phase, one pulse $\pi$ out of phase, and only two pulses interfering, respectively. A simple Gaussian beam model is used to generate the patterns for comparison; these are shown in Figs. 1(c), 1(e) and 1(g) for the same conditions as those above. The replica focus allows for phase measurement of the incident four beam without the tracer impinging the sample during the TFWM and 2DFT experiments.

Samples are epitaxially grown and consist of 4 periods of GaAs QWs surrounded by $Al_{0.3}Ga_{0.7}As$ barriers. Both the QWs and barriers are 10 nm thick. For the experiments the samples are attached to sapphire disks and thinned. Experiments are performed at approximately 5 K. Linear absorbance is shown in Fig. 2(a), where $X_{hh}$ and $X_{lh}$ peaks are the heavy- and light-hole excitons, respectively.

Rephasing 2DFT measurements are performed by scanning pulse A away from the sample (backward in time) in uniform steps while measuring the complex TFWM field. A Fourier transform is performed along the direction of the scanned time axis. The complex field is captured by SI with the reference. Figures 2(b) – 2(d) show an example interference term, retrieved TFWM spectrum (red) and spectral phase, respectively. The non-heterodyned TFWM spectrum (black) shows good agreement. The polarizations of these examples are co-linear, that is XXXX, where the first three X's denote the polarization state of the pump beams and the final that of the emission. For cross-linear polarized measurements the excitation is YXXY.

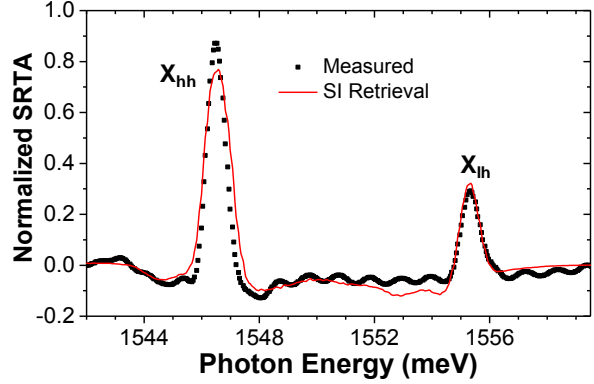

Fig. 3. SRTA measurement accompanied by the phase rotated complex TFWM spectrum.

## 4. Results and Discussion

### 4.1 Phase retrieval

Figure 3 shows the measured SRTA (black dots) and the phase rotated complex TFWM signal (red line). These are measured under the same initial excitation conditions as the start position of the 2DFT scans shown in the next subsection, i.e. $T = 200$ fs, co-linear polarization and an exciton density of about $3 \times 10^{10}$ cm$^{-2}$ per layer. Under these conditions the comparison SRTA measurement normally performed well. The obtained phase for this is $\phi_{SRTA} = 0.99\,\pi$, with an error of $\pm 0.02\,\pi$, determined by several repetitions of the measurement. Despite this low error value, using the auxiliary SRTA measurement can be problematic:

- The pump and probe pulses do not correctly mimic the three-beam TFWM signal.
- At low excitation densities the reproducibility of the SRTA is poor.
- SRTA is not applicable to cross-polarized 2DFT spectra.
- Similarly SRTA is not applicable to projections yielding two-quantum information.

The first two points are technical and can potentially be overcome. The latter two however, is a fundamental restriction to the versatility of 2DFT spectroscopy.

The phase offsets outlined in Eq. (9) are required to capture the global phase using the all-optical method proposed in the previous section. It should be made clear that the phase offsets are very sensitive to the time delays between the pulses [2]. Therefore, field correlations between all the pulse pairs must be performed at the replica focus. Figure 4(a) shows the field correlation between pulses A and B. The $\tau = 0$ position on the delay stage is found by fitting the pattern with a Gaussian envelope on a sinusoidal oscillation. The error associated with fitting is up to 1 fs, which is potentially a 0.04 π error at 800 nm. This is repeated for each pair in the box geometry. For correct phase retrieval of the $S_I(\omega_\tau, T, \omega_t)$ 2DFT scan, the initial delays $\tau_0$ and $t_0$ must be known precisely. The former is set using the field correlation method, where as the latter is determined though SI, discussed below.

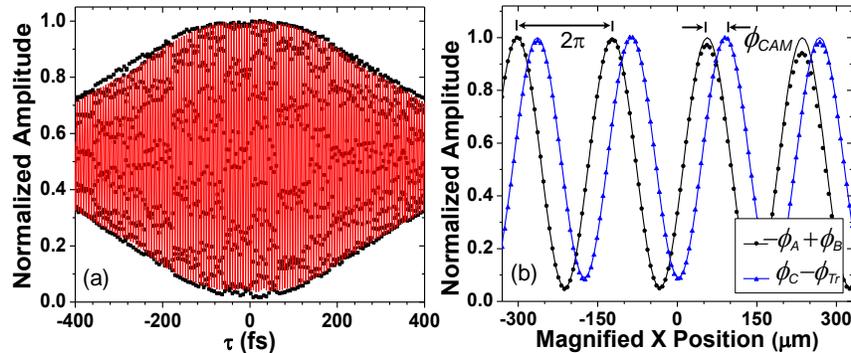

Fig. 4. (a) Field correlation between beams A and B: measured points are black dots which are fit with a sine wave and a Gaussian envelope. (b) Lineouts of patterns between pulses A and B (black dots), and pulses C and Tr (blue dots). Both are fit with a sine wave to find the phase.

Once the time delays are set correctly, the phase offsets between the incident pumps and tracer pulses are extracted from the interference patterns recorded with the camera image. Examples of two-dimensional patterns are shown in Figs. 1(b) – 1(e). However, it is easier to fit one-dimensional interference patterns between pairs of pulses; see Figs. 1(f) and 1(g). Vertical integration of the center section of the camera images for the pairs yields one-dimensional "*lineouts*" that can be fit with a sine wave. Figure 4(b) shows the *lineouts* for $-\phi_A + \phi_B$ and $\phi_C - \phi_T$ from which $\phi_{CAM}$ is determined, giving a value of about 0.50 ± 0.04 π. The error here is associated with fitting the interference pattern. Additional errors may occur if the pulses in the box are not perfect copies of one another, leading to imperfect interference patterns at the replica focus. Also, note that improved phase retrieval can be obtained by applying small phase shifts of the interference patterns, captured before and after locking the feedback loop.

Spectral phases $\phi_{TR}(\omega)$, between the tracer and reference, and $\phi_{SR}(\omega)$, between the signal and reference are measured though SI. From the spectral phases the phase offsets $\phi_{SR}$ and $\phi_{TR}$ can be determined by comparison of the spectral phase with and without an offset. Figure 2(d) shows the value for $\phi_{SR} \sim 1.65$ π ± 0.04 π, extracted from the center plateau between the $X_{hh}$ and $X_{lh}$ resonances. Details of the tracer-reference SI are shown in Fig. 5, where the interfeometric term is presented in panel 5(a), with its inverse Fourier transform in panel 5(b). In the time domain, the zero-time contributions (red) can be truncated leaving only the contribution due to spectral interferometry with the tracer (black). As can be seen, the time delay between the tracer and reference is approximately 15 ps. This is $t_0$, which is applied in the 2DFT analysis to prevent phase roll along the emission axis. The accuracy in retrieving $t_0$ is about 5 fs, which is suitable since the spectral region of interest is less than half the laser bandwidth. Figures 5(c) and 5(d) show the retrieved spectrum and spectral phase of the tracer, respectively. In the latter, the experimental phase (black dots) is modeled using a vector sum of the incident radiation and the polarization of the sample, where the polarization is obtained from the linear absorption. The modeled phase has two discontinuities, one at each resonance, where the reradiated light is in anti-phase with the incident light. The $X_{hh}$ peak dominates the phase structure, because it has the larger dipole moment. In addition, the modeling requires a quadratic spectral phase due to slight differences in the dispersion between the tracer and reference arms of the interferometer. From the comparison a value of $\phi_{TR} = 0.11\,\pi \pm 0.05\,\pi$ is used for the retrieved phase offset. The combined phase offsets from Figs. 2(d), 4(b) and 5(d) give a global phase of the signal of about $\phi_{P^{(3)}} = 1.04\,\pi$, with a combined error of up to ±0.08 π. This value can be improved by averaging several 2DFT spectra. The all-optical phase compares favorably with the phase retrieval by SRTA method, and moreover it can be used in cases where the SRTA is not applicable.

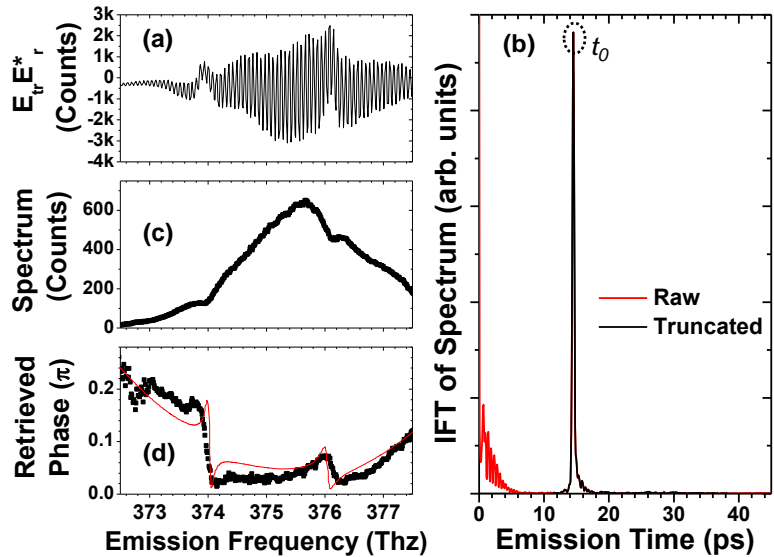

Fig. 5. The tracer-reference SI measurement: (a) the interferometric term, (b) the inverse Fourier transform of the interferogram (with and without truncation), (c) the retrieved spectrum of the tracer and (d) the retrieved spectral phase (black dots) with its modeled phase (red line).

*4.2 Co-linearly Polarized 2DFT Measurements*

Co-linear polarized 2DFT spectra with $T = 200$ fs are shown in Fig. 6 with real and imaginary parts in the left and right panels respectively. The top panels 6(a) and 6(b) are *phased* using the SRTA method and the bottom panels 6(c) and 6(d) are the same data *phased* using the all-optical method. The agreement between the spectral lineshapes is very good. In addition the lineshape matches well to results recently shown in a polarization study of a 10 period GaAs MQW sample [24].

The 2DFT spectra show two diagonal features corresponding the $X_{hh}$ and $X_{lh}$ populations and coupling between both these states with off-diagonal peaks. (Note that the absorption axis is negative, and hence the diagonal slopes downward.) Elongation is also observed due to inhomogeneous broadening of the states. Strong many-body interactions dominate the co-linear polarized spectrum; consequently the diagonal lineshapes of the real part are dispersive. The $X_{hh}$ peak dominates the spectrum because of its stronger oscillator strength. Of the two off-diagonal coupling peaks the $X_{lh\text{-}hh}$ (bottom left) is the strongest, due to the strong coherent many-body interactions in the $X_{hh}$ and $X_{lh}$ system [23]. The imaginary part of the spectrum shows absorptive lineshapes for all four peaks. Additionally, the heavy-hole biexciton ($2X_{hh}$) can be clearly seen as a negative dip. This is expected since the $2X_{hh}$ is an excited state from which a coherent oscillation is generated.

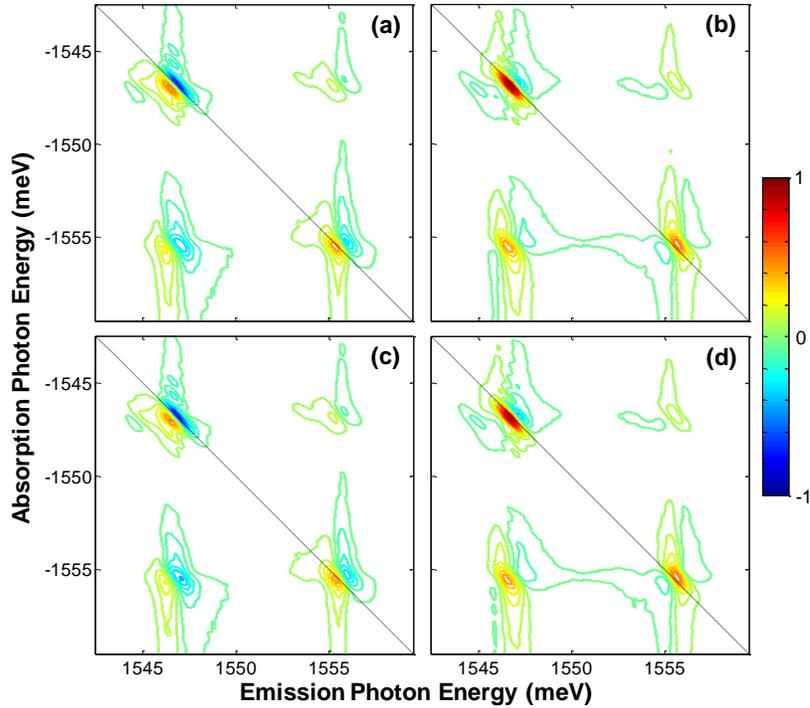

Fig. 6. Co-linear polarized $S_I(\omega_\tau,T,\omega_t)$ 2DFT spectra with $T = 200$ fs. Panels (a) and (b) show the real and imaginary part phased by the SRTA technique. Panels (c) and (d) are the same but phased using the all-optical method.

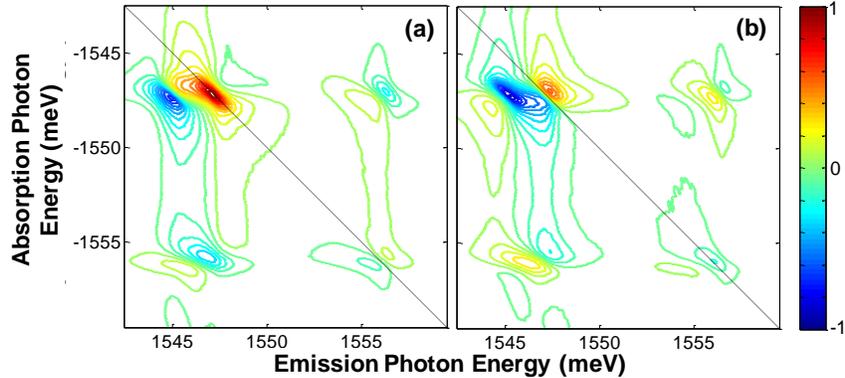

Fig. 7. (a) Real and (b) imaginary parts of the cross-linear 2DFT spectrum for $T = 0$ fs.

*4.3 Cross-linearly Polarized 2DFT Measurements*

Figure 7 shows the real 7(a) and imaginary 7(b) parts of the cross-linear polarized 2DFT spectrum taken for $T = 0$ fs. (Due to a significantly weaker TFWM signal for cross-polarization excitation $T = 200$ fs is harder to obtain.) The polarization study of Zhang *et al.* could not capture the correct global phase using SRTA and only showed the amplitude spectrum with horizontal elongation due to biexciton peaks on all four features [24].

The real part of the spectrum shows absorptive diagonal features, as would be expected due to the suppression of many-body interactions for these polarization conditions. Additionally, the $2X_{hh}$ peak is stronger and once again a negative dip. Cross-polarization enhances the relative strength of the biexciton peaks due to selection rules. The real part of the spectrum shows similarities to calculations performed using a one-dimensional microscopic site-based tight-binding model [28].

## 5. Conclusion

We have demonstrated an all-optical method for obtaining the global phase for a 2DFT spectroscopy. The method requires measuring the phase difference between the input pump beams and the tracer beam, and performing SI between the tracer and a phase-stabilized reference beam used for heterodyne detection. The method has been contrasted with the SRTA technique for correctly *phasing* the 2DFT spectra, showing promising results. An additional demonstration has been given where SRTA is not applicable, i.e. with the first two pump pulses cross-polarized. This method has been presented only for the common rephasing $S_I(\omega_\tau, T, \omega_t)$ projection of 2DFT spectroscopy, but is applicable to both the $S_{II}$ and $S_{III}$ 2DFT techniques, which correspond to the phase-matching conditions $k_{SII} = k_A - k_B + k_C$ and $k_{SIII} = k_A + k_B - k_C$, respectively. A similar technique has been proposed for infrared 2DFT spectroscopy by Backus *et al* [29].

**Acknowledgements**

The authors wish to thank David Jonas, Ralph Jimenez and Tianhao Zhang for useful discussions, and Richard Mirin for providing samples. This work was supported by the NSF and the Chemical Sciences, Geosciences, and Biosciences Division Office of Basic Energy Sciences, U.S. Department of Energy.